\documentclass[twocolumn,showpacs,floats,floatfix,superscriptaddress,aps,pra]{revtex4-1}
\usepackage{amsfonts}
\usepackage{amssymb}
\usepackage{amsmath}
\usepackage{calc}
\usepackage{graphicx}
\usepackage{bm}

\usepackage[normalem]{ulem}
\usepackage{amsmath,amssymb}
\usepackage{multirow}
\usepackage{xcolor,soul}
\usepackage{srcltx}
\usepackage{hyperref,graphicx}

\def\be{ \begin{equation}}
\def\ee{ \end{equation}}
\def\bea{ \begin{eqnarray}}
\def\eea{ \end{eqnarray}}
\def\bse{ \begin{subequations}}
\def\ese{ \end{subequations}}
\def\bc{ \begin{center}}
\def\ec{ \end{center}}

\def\ket#1{\vert #1 \rangle}
\def\bra#1{\langle #1 \vert}
\def\braket#1#2{\langle #1 \vert #2 \rangle}

\def\at{\text{at}}
\def\bos{\text{bos}}

\begin{document}

\author{Boyan T. Torosov}
\altaffiliation{Permanent address: Institute of Solid State Physics, Bulgarian Academy of Sciences, 72 Tsarigradsko chauss\'{e}e, 1784 Sofia, Bulgaria\\
e-mail: torosov@phys.uni-sofia.bg}
\affiliation{Dipartimento di Fisica, Politecnico di Milano and Istituto di Fotonica e Nanotecnologie del Consiglio Nazionale delle Ricerche, Piazza L. da Vinci 32, I-20133 Milano, Italy}
\author{Stefano Longhi}
\affiliation{Dipartimento di Fisica, Politecnico di Milano and Istituto di Fotonica e Nanotecnologie del Consiglio Nazionale delle Ricerche, Piazza L. da Vinci 32, I-20133 Milano, Italy}
\author{Giuseppe Della Valle}
\affiliation{Dipartimento di Fisica, Politecnico di Milano and Istituto di Fotonica e Nanotecnologie del Consiglio Nazionale delle Ricerche, Piazza L. da Vinci 32, I-20133 Milano, Italy}
\title{Mixed Rabi Jaynes-Cummings model of a three-level atom interacting with two quantized fields}
\date{\today}

\begin{abstract}
The quantum Rabi model  describes the ultrastrong interaction of a two-level atom coupled to a single quantized bosonic mode. As compared to the Jaynes-Cummings model, in the Rabi model the absorption and emission processes do not need to satisfy energy conservation and the usual rotating wave approximation (RWA) breaks down. As a result, the atom-field dynamics in the Hilbert space splits into two independent parity chains, exhibiting a collapse-revival pattern and exact periodic dynamics in the limit of degenerate atomic levels. Here we introduce a mixed Rabi Jaynes-Cummings model by considering a three-level atom interacting with two quantized bosonic fields, in which the RWA is made for one transition (with a weak atom-field coupling) but not for the other one (with an ultrastrong atom-field coupling). As a result, we show that the field in the weak coupled atomic transition can be used as a tool to control the atom-field dynamics of the other (strong coupled) transition, thus realizing an effective two-level quantum Rabi model with a controllable field. In particular, a periodic temporal dynamics of the atom-field state can be realized by appropriate tuning of the weak control field, even for non-degenerate atomic levels. A photonic simulator of the mixed Rabi Jaynes-Cummings model, based on light transport in evanescently-coupled optical waveguide lattices, is also briefly discussed.
\end{abstract}

\maketitle

\section{Introduction}

The well-known quantum Rabi model \cite{Rabi1a,Rabi1b,Rabi1c,Rabi1d}, describing a two-level atom coupled to a quantum harmonic oscillator, continues to produce rich and surprising physics, with plenty of applications in a variety of physical systems. The quantum Rabi model has been applied to numerous experimental systems in quantum optics or condensed matter, such as cavity quantum electrodynamics (QED) \cite{QEDa,QEDb,QEDc}, quantum dots \cite{dots}, superconducting qubits \cite{SCa,SCb} and trapped ions \cite{ionsa,ionsb}. In most cases, when the external field is weak enough \cite{note}, the rotating wave approximation (RWA) is applied and in such way the famous Jaynes-Cummings model is obtained \cite{JC}. However, in recent years, new regimes have been explored \cite{Wolf,Tunneling,USCa,USCb,USCc,USCd,USCe,DSCa,DSCb}, in which the effect of counter-rotating terms cannot be neglected. Such regimes are the ultrastrong coupling of light-matter interactions \cite{USCa,USCb,USCc,USCd,USCe} and the deep strong coupling (DSC) \cite{DSCa,DSCb}. In the DSC regime, the absorption and emission processes do not need to satisfy energy conservation and the atom-field dynamics is more involved and splits into two parity chains in Hilbert space. As a result, the atom-field state undergoes revival and collapse dynamics in Hilbert space \cite{DSCa}. Remarkably, in the limit of degenerate atomic levels the dynamics becomes exactly periodic \cite{DSCa}. 
Recent works have shown that the quantum Rabi model can be simulated by using light transport in engineered waveguide superlattices \cite{Longhia,Longhib, Rodriguez}. This could allow the DSC regime, which is hard to access experimentally in cavity QED, to be successfully simulated in other physical contexts. In spite of the vast research in this area and the relative simplicity of the Rabi model, its integrability has been proven just recently \cite{Braaka,Braakb,Braakc}. 

In the past few decades, several theoretical and experimental works have shown that many interesting coherent phenomena can be observed when more than two atomic levels are involved in the dynamics. In particular, the three level system exhibits a plethora of coherent phenomena such as two-photon coherence \cite{add3}, resonance Raman scattering \cite{add4}, double resonance process \cite{add5}, three-level super radiance \cite{add7} and quantum jumps \cite{add10}, to mention a few. The quantum dynamics of a three-level atom interacting with two resonant or near resonant modes of a steady field has been generally studied by assuming either quantized fields, using a dressed-state formalism within the RWA (extended Jaynes-Cummings models \cite{NstatesRabia,NstatesRabib,NstatesRabic}), or  the Floquet approach for classical fields \cite{uff1,uff2,uff3,uff4}. In such previous works, the strong coupling regime was considered solely in the semiclassical limit, i.e. for classical fields, where a Floquet analysis of the underlying time-periodic equations for the atomic population amplitudes can be employed for sinusoidal external fields \cite{uff0}. However, for quantized fields the ultrastrong coupling regime in the three-level atomic system, which breaks the RWA, was not investigated.

In this paper, we study in details a three-level atomic system interacting with two quantized fields, where one field is near-resonant and weakly coupled with one atomic transition whereas the other field is strongly coupled to the other atomic transition. Such a quantum model can be refereed to as mixed Rabi Jaynes-Cummings model, because the RWA can be applied to one transition, but not to the other one. We derive the coupled differential equations, describing the temporal evolution of the quantum system in Hilbert space, and  show that the weak-coupling filed  can be used as a tool to control the dynamics in Hilbert space of the atom-field state for the other transition. In particular, the weak control field can be tuned to realize exact periodic of the atom-field state even if the strongly-coupled atomic levels are not degenerate. A possible physical implementation of the mixed Rabi Jaynes-Cummings model, using arrays of coupled optical waveguides with engineered coupling constants, is also briefly discussed.


\begin{figure}[tb]
\includegraphics[width=9cm]{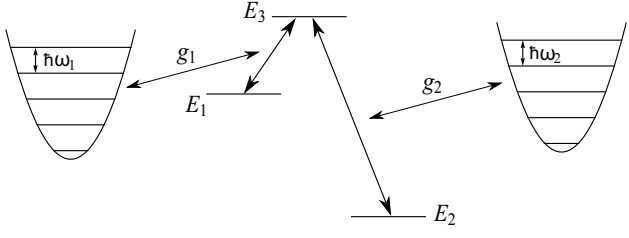}
\caption{Schematic of a three-level system interacting with two bosonic modes with $n$ quanta at frequency $\omega_1$ and $k$ quanta at frequency $\omega_2$.}
\label{system}
\end{figure}

\section{The model}
We consider a three-level quantum system, with atomic states $\ket{1}_\at$, $\ket{2}_\at$, and $\ket{3}_\at$, interacting with two bosonic (e.g. electromagnetic) fields of states $\ket{n,k}_\bos$, where $n$ and $k$ are the number of bosons in the two fields (Fig.~\ref{system}). Such system is described by the Hamiltonian
\begin{align}\label{HRabi}
&H=\sum_{i=1}^{3}E_i\sigma_{ii}+\hbar\omega_1 a_1^\dagger a_1+\hbar\omega_2 a_2^\dagger a_2 \\ \notag
&+\hbar g_1(\sigma_{13}+\sigma_{31})(a_1+a_1^\dagger)+\hbar g_2(\sigma_{23}+\sigma_{32})(a_2+a_2^\dagger),
\end{align}
where $\omega_{1,2}$ are the frequencies of the fields that are responsible for the $\ket{1}_\at-\ket{3}_\at$ and $\ket{2}_\at-\ket{3}_\at$ atomic transitions, respectively. The coupling strengths are parameterized by $g_1$ and $g_2$, $a_{1,2}^\dagger$ and $a_{1,2}$ are the creation and annihilation operators for the two modes and $\sigma_{ij}=\ket{i}_{\at} {\,}_{\at}\bra{j}$. We now assume $\omega_1\ll\omega_2$ and a resonant and weak coupling for the $\ket{2}_\at-\ket{3}_\at$ transition, hence $\hbar\omega_2 = E_3 - E_2$ and $\hbar g_2\sqrt{k}\ll E_3 - E_2$. Under these conditions we can apply the RWA for this transition, discarding the counter-rotating terms $\sigma_{32}a_2^{\dagger}$ and $\sigma_{23}a_2$, and the Hamiltonian now reads
\begin{figure}[tb]
\includegraphics[width=9cm]{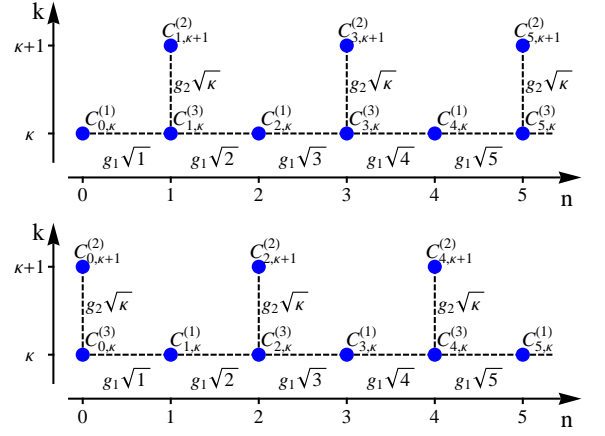}
\caption{Coupling scheme after RWA for the $\ket{2}_\at-\ket{3}_\at$ transition. Depending on the initial condition, the chain contains only (top) even number or (bottom) odd number of bosons in the first mode when the atom is in state $\ket{1}_\at$.}
\label{grid}
\end{figure}
\begin{align}\label{HRabiRWA}
&H=\sum_{i=1}^{3}E_i\sigma_{ii}+\hbar\omega_1 a_1^\dagger a_1+\hbar\omega_2 a_2^\dagger a_2 \\ \notag
&+\hbar g_1(\sigma_{13}+\sigma_{31})(a_1+a_1^\dagger)+\hbar g_2(\sigma_{32}a_2+\sigma_{23}a_2^\dagger).
\end{align}
The model described by the Hamiltonian~\eqref{HRabiRWA} is a mixed Rabi and Jaynes-Cummings model, because the RWA is performed for one of the two transitions (like in standard Jaynes-Cummings model) but not for the other one (like in the quantum Rabi model). 
To study the exact temporal evolution of the atom-field state $\ket{\Psi(t)}$ in the mixed Rabi Jaynes-Cummings model, let us expand the state vector of the system as
\begin{align}\notag
&\ket{\Psi(t)}=\\ \label{Psi}
&\sum_{n,k}\left[ C_{n,k}^{(1)}(t)\ket{1}_\at+C_{n,k}^{(2)}(t)\ket{2}_\at+C_{n,k}^{(3)}(t)\ket{3}_\at \right]\ket{n,k}_\bos\, ,
\end{align}
where $C_{n,k}^{(l)}$ is the probability amplitude to have $(n,k)$ bosons in the two fields and the atom in level $\ket{l}_\at$. Substitution of the ansatz \eqref{Psi} into the Schr\"{o}dinger equation
\be
i\hbar\partial_t\ket{\Psi(t)}=H\ket{\Psi(t)}\, ,
\ee
yields the following coupled differential equations for the probability amplitudes,
\bse\label{coupling eqs}
\begin{align}
i\hbar\dot{C}_{n,k}^{(1)}&= E_1 C_{n,k}^{(1)}+\left( n\hbar\omega_1 + k\hbar\omega_2 \right)  C_{n,k}^{(1)} \\ \notag
&+\hbar g_1\sqrt{n+1}C_{n+1,k}^{(3)}+\hbar g_1\sqrt{n}C_{n-1,k}^{(3)}\\
i\hbar\dot{C}_{n,k}^{(2)}&= E_2 C_{n,k}^{(2)} + \left( n\hbar\omega_1 + k\hbar\omega_2 \right)  C_{n,k}^{(2)}\\ \notag
&+\hbar g_2\sqrt{k}C_{n,k-1}^{(3)}\\
i\hbar\dot{C}_{n,k}^{(3)}&=E_3 C_{n,k}^{(3)}+\left( n\hbar\omega_1 + k\hbar\omega_2 \right)  C_{n,k}^{(3)}\\ \notag
&+\hbar g_1\sqrt{n+1}C_{n+1,k}^{(1)}+\hbar g_1\sqrt{n}C_{n-1,k}^{(1)}\\ \notag
&+\hbar g_2\sqrt{k+1}C_{n,k+1}^{(2)}\, .
\end{align}
\ese	
The coupling between the amplitudes can be visualized as two sets of uncoupled chains, where each lattice site corresponds to a different state of the atom-field system. Depending on the initial condition, one of the two sets is realised and the other one is irrelevant. The chains in each set are uncoupled in the $k$ direction, because of the RWA, while in the $n$ direction they are semi-infinite with a gradient in the boson number. This is depicted in Fig.~\ref{grid}, where we show the two coupling schemes, for a particular value of the second-mode boson number $k=\kappa$. The top part of the picture shows the coupling scheme with only even number $n$ of bosons in the first mode when the atom is in state $\ket{1}_\at$. On the opposite, the bottom part contains only odd number of bosons $n$ when the atom is in $\ket{1}_\at$. In what follows, we will assume that the initial condition is $\ket{\Psi(0)}=\ket{1}_\at\ket{0,\kappa}_\bos$ and hence the top part of Fig.~\ref{grid} is relevant. 
A more suitable basis for the vector state of the system is $[\ket{b_1},\ket{b_2},\ket{b_3}, \ket{b_4},\ket{b_5},\ket{b_6},\ldots]$, where
\begin{eqnarray}
\ket{b_1} &=& \ket{1}_\at\ket{0,\kappa}_\bos,\nonumber \\
\ket{b_2} &=& \ket{3}_\at\ket{1,\kappa}_\bos,\nonumber \\
\ket{b_3} &=& \ket{2}_\at\ket{1,\kappa+1}_\bos,\nonumber \\
\ket{b_4} &=& \ket{1}_\at\ket{2,\kappa}_\bos,\\
\ket{b_5} &=& \ket{3}_\at\ket{3,\kappa}_\bos,\nonumber \\
\ket{b_6} &=& \ket{2}_\at\ket{3,\kappa+1}_\bos,\nonumber \\
\vdots && \nonumber
\end{eqnarray}
In this basis, the Hamiltonian can be represented in the following matrix form:
\begin{widetext}
\be\label{Hinf}
H = \left[\begin{array}{ccccccc}
E_1    & \hbar g_1\sqrt{1}       & 0                               & 0                  & 0                  & 0      														 & 0\\
\hbar g_1\sqrt{1}& E_3+\hbar\omega_1 & \hbar g_2\sqrt{\kappa}                 & \hbar g_1\sqrt{2}        & 0                  & 0      														 & 0 \\
0      & \hbar g_2\sqrt{\kappa}       & E_3+\hbar\omega_1 & 0                  & 0                  & 0      														 & 0 \\
0      & \hbar g_1\sqrt{2}       & 0                               & E_1+2\hbar\omega_1 & g_1\sqrt{3}        & 0      														 & 0 \\
0      & 0                 & 0                               & \hbar g_1\sqrt{3}        & E_3+3\hbar\omega_1 & \hbar g_2\sqrt{\kappa}												 & \hbar g_1\sqrt{4} \\
0      & 0                 & 0                               & 0                  & \hbar g_2\sqrt{\kappa}        & E_3+3\hbar\omega_1    & 0\\
0 		 & 0           			 & 0                		           & 0           		    & \hbar g_1\sqrt{4}		     & 0  		 														 & \ddots 
\end{array}\right] ,
\ee
\end{widetext}
where we have discarded the common diagonal term $\kappa\hbar\omega_2$ and we have used the resonance condition $E_3=E_2+\hbar\omega_2$. 
\begin{figure}[tb]
\includegraphics[width=15cm]{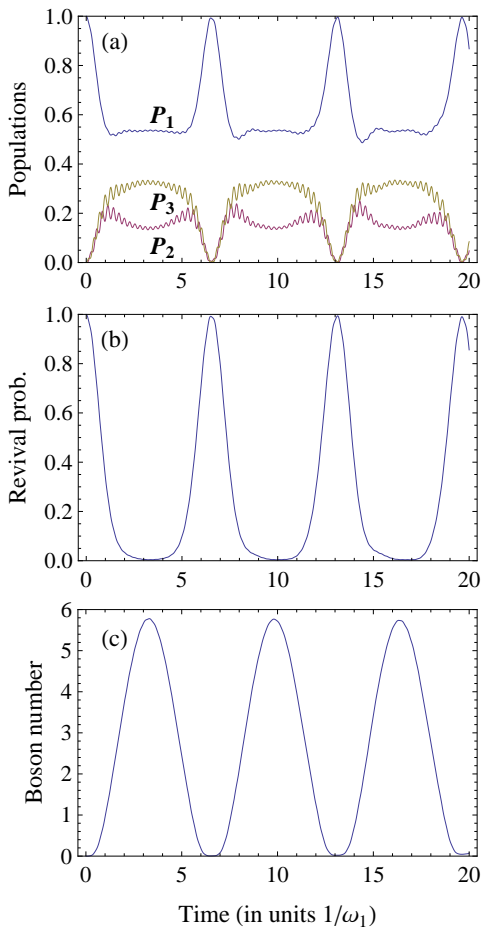}
\caption{Time evolution of (a) the populations of states $\ket{1}_\at$, $\ket{2}_\at$ and $\ket{3}_\at$, (b) revival probability $\left|\bra{\Psi(0)}\ket{\Psi(t)}\right|^2$ and (c) boson number in the first mode. The energies of the bare states are $E_1=90\hbar\omega_1$, $E_2=0$, $E_3=100\hbar\omega_1$. The coupling strengths are $g_1=1.5\omega_1$ and $\hbar g_2\sqrt{\kappa}=E_3-E_1$ and the mode frequencies are $\omega_2=100\omega_1$.}
\label{populations}
\end{figure}
\begin{figure}[tb]
\includegraphics[width=15cm]{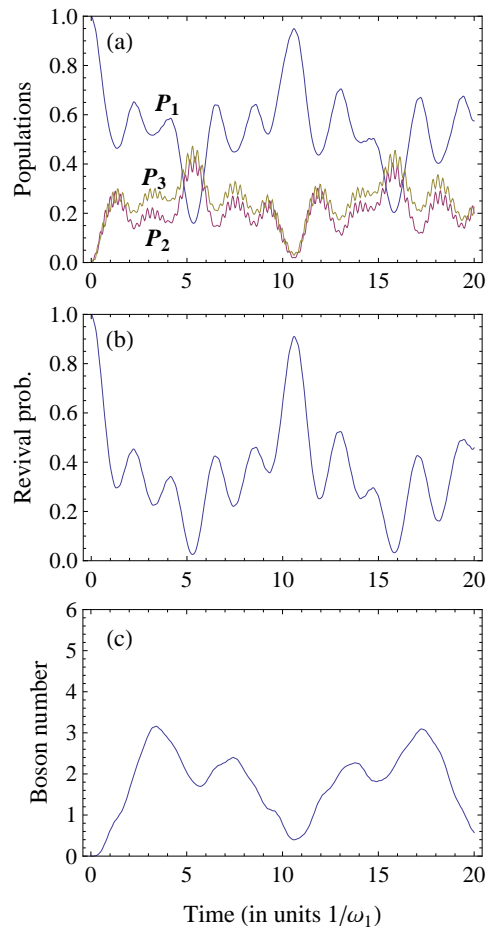}
\caption{Same as Fig. \ref{populations}, but for $\hbar g_2\sqrt{\kappa}=E_3-E_1+2\hbar\omega_1$.}
\label{populations2}
\end{figure}
It should be noted that the type of Hamiltonian matrix given by Eq.~\eqref{Hinf}, obtained by projection of the full quantized Hamiltonian in the atom-field basis $  \ket{l}_\at\,   \ket{n,k}_\bos\,$ ($l=1,2,3$), has a similar structure of the Floquet matrix that one would obtain in the semiclassical limit, i.e. when the strong field at frequency $\omega_1$ is assumed to be a classical field (coherent state). In the semiclassical analysis, the equations of the atomic field amplitudes without the RWA for the transition $\ket{2}_\at-\ket{3}_\at$ are coupled by the oscillating field at frequency $\omega_1$, i.e. they form a set of non-autonomous equations with periodic coefficients. The semiclassical time-periodic Hamiltonian of the atomic amplitude probabilities can be cast into an equivalent infinite-dimensional time-independent matrix Hamiltonian by standard Floquet approach \cite{uff0}. The resulting Floquet matrix has a form similar to Eq.~\eqref{Hinf}.  The main difference between the classical and quantized field descriptions is that in the latter case the couplings, described by the two second main diagonals of Eq.~\eqref{Hinf}, depend on the photon number.
 Since the spectrum of the matrix $H$ is generally not equally spaced, the atom-field dynamics is generally aperiodic, regardless the field at frequency $\omega_1$ is considered classical or quantized. In particular, if the control field at frequency $\omega_2$ is switched off ($g_2=0$), the Hamiltonian $H$ reduces to that of the two-level quantum Rabi. In this case it is known that the spectrum of  $H$ is not equally-spaced, unless the degenerate level limit $E_3-E_1 \rightarrow 0$ is considered \cite{DSCa}. In the semiclassical limit, the lack of periodicity can be explained on the basis of Floquet theory of the strongly-driven two-level atomic transition, in which two generally incommensurate frequencies are involved (the frequency $\omega_1$ of the external field and the difference of the quasi energies; see, for instance, \cite{driven2}). 
 To calculate the energy spectrum of $H$ and the atom-field dynamics in the general case $g_2 \neq 0$, one has to resort  rather generally to a numerical analysis.
However, some interesting physics can be found if one takes a different approach. We note that the coupling strength $g_2\sqrt{\kappa}$ splits the common energy $E_3+\hbar \omega_1 n = E_2+\hbar\omega_2 +\hbar \omega_1 n$ of the states $\ket{3}_\at\ket{n,\kappa}_\bos$ and $\ket{2}_\at\ket{n,\kappa+1}_\bos$  into an Autler-Townes doublet, and in such way a dressed picture is obtained, where the energies of the two dressed states are equal to $E_3+\hbar\omega_1 n\pm \hbar g_2\sqrt{\kappa}$. If now we set $\hbar g_2\sqrt{\kappa}=E_3-E_1$ we obtain for the lower energy of the dressed states $E_3+\hbar\omega_1 n - \hbar g_2\sqrt{\kappa} = E_1+\hbar\omega_1 n$. In such a way, for $g_2\sqrt{\kappa}\gg g_1\sqrt{n}$, we obtain a one-dimensional coupled chain with a boson number gradient. This corresponds to the limit of degenerate qubit levels of the two-level quantum Rabi model, which was described in \cite{Longhia,Longhib}. This system is characterized with a strictly periodic behavior of the populations and the boson number. We illustrate this on Fig.~\ref{populations}, where we plot the time evolution of the populations of the three levels, the revival probability $\left|\braket{\Psi(0)}{\Psi(t)}\right|^2$, and the boson number as a function of time. The initial condition is $\ket{\Psi(0)}=\ket{1}_\at\ket{0,\kappa}_\bos$. The figure clearly reveals a periodic dynamics with a period of approximately $2\pi/\omega_1$. We also notice that the mean boson number of the first mode never obtains too large values, which justifies the use of the approximation $g_2\sqrt{\kappa}\gg g_1\sqrt{n}$. On the contrary, if the coupling strength $\hbar g_2\sqrt{k}$ is not tuned to $E_3-E_1$, we will get a non-periodic behavior of the dynamics. This is shown in Fig.~\ref{populations2}.

\section{Conclusions and discussion}

In this paper we have studied theoretically a mixed Rabi Jaynes-Cummings model of  three-level atom interacting with two quantized bosonic modes, for which the RWA can be applied for one transitions solely. We have derived the differential equations, describing the evolution of the system in Hilbert space, and shown that the weak-coupled bosonic mode can be exploited to control the dynamics of the atom-field of the strong-coupled transition. In particular a transition from non-periodic to near-periodic behavior of the evolution has been found by appropriate tuning of the control weak field. Such a transition has been related to the spectrum of the  two-level quantum Rabi Hamiltonian, which shows an equally-spaced energy ladder structure in the limit of degenerate energy levels. As a final comment, let us briefly mention that the mixed Rabi Jaynes-Cummings model introduced in the present work can be simulated in an optical setting, following an approach similar to the one recently proposed and demonstrated in Refs.~\cite{Longhia,Longhib}. 
The coupling scheme shown in Fig.~\ref{grid} can be implemented in classical optics by using evanescently-coupled dielectric waveguides. For this aim, one should notice that Eqs.~\eqref{coupling eqs} are analogous to the coupled-mode equations describing light transport in a semi-infinite one-dimensional photonic lattice in the tight-binding approximation with additional waveguides, which couple to only the even-number waveguides from the chain. The lattice shows a superimposed transverse index gradient and a non-uniform coupling constant between adjacent waveguides. The coupling constants are controlled by the distance $d_n$ between the waveguides and the index gradient $\omega_1$ is given by $\omega_1=2\pi n_s a/(R\lambda)$, where $n_s$ is the refractive index of the substrate, $\lambda$ is the wavelength of light, $R$ is the bending radius of curvature and $a$ is the horizontal spacing of the waveguides (see, for instance \cite{Longhi-review,Garanovich}). The coupling strength $J_n=\hbar g_1\sqrt{n}$ between the neighboring waveguides is given to an excellent accuracy by the exponential law $J_n=\chi \exp(-\alpha d_n)$, where $\chi$ and $\alpha$ are some constants which depend on the waveguide fabrication parameters and can be experimentally determined.
 
\acknowledgments

This work was supported by the Fondazione Cariplo (Grant No. 2011-0338).



\begin{thebibliography}{99}

\bibitem{Rabi1a} I. I. Rabi, Phys. Rev. 49 (1936) 324.
\bibitem{Rabi1b} I. I. Rabi, Phys. Rev. 51 (1937) 652.
\bibitem{Rabi1c} S. Haroche and J.-M. Raimond, Exploring the Quantum: Atoms, Cavities, and Photons, Oxford University Press, Oxford, 2006.
\bibitem{Rabi1d} C. C. Gerry and P. L. Knight, Introductory Quantum Optics, Cambridge University Press, Cambridge, England, 2004.

\bibitem{QEDa} S. Haroche and D. Kleppner, Phys. Today 42 (1989) 24.
\bibitem{QEDb} B. W. Shore and P. L. Knight, J. Mod. Opt. 40 (1993) 1195.
\bibitem{QEDc} R. Miller, T. E. Northup, K. M. Birnbaum, A. Boca, A. D. Boozer and H. J. Kimble, J. Phys. B: At. Mol. Opt. Phys. 38 (2005) S551.

\bibitem{dots} J. P. Reithmaier, G. Sek, A. L\"{o}ffler, C. Hofmann, S. Kuhn, S. Reitzenstein, L. V. Keldysh, V. D. Kulakovskii,
T. L. Reinecke, and A. Forchel, Nature 432 (2004) 197.

\bibitem{SCa} A. Wallraff, D. I. Schuster, A. Blais, L. Frunzio, R.- S. Huang, J. Majer, S. Kumar, S. M. Girvin and R. J. Schoelkopf, Nature 431 (2004) 162.
\bibitem{SCb} D. I. Schuster, A. A. Houck, J. A. Schreier, A. Wallraff, J. M. Gambetta, A. Blais, L. Frunzio, J. Majer, B. R. Johnson, M. H. Devoret, S. M. Girvin, and R. J. Schoelkopf, Nature 445 (2007) 515.

\bibitem{ionsa} D. F. V. James,  Appl. Phys. B 66 (1998) 181.
\bibitem{ionsb} D. Leibfried, R. Blatt, C. Monroe, and D. Wineland, Rev. Mod. Phys. 75 (2003) 281.


\bibitem{note} Weak enough means that the Rabi frequency should be much less than the transition frequency.

\bibitem{JC} E. T. Jaynes and F. W. Cummings, Proc. IEEE 51 (1963) 89.


\bibitem{Wolf} F. A. Wolf, F. Vallone, G. Romero, M. Kollar, E. Solano, and D. Braak, Phys. Rev. A 87 (2013) 023835.

\bibitem{Tunneling} E. K. Irish and J. Gea-Banacloche, Phys. Rev. B 89 (2014) 085421.

\bibitem{USCa} C. Ciuti, G. Bastard, and I. Carusotto, Phys. Rev. B 72 (2005) 115303.
\bibitem{USCb} J. Bourassa, J. M. Gambetta, A. A. Abdumalikov, Jr., O. Astafiev, Y. Nakamura, and A. Blais, Phys. Rev. A 80 (2009) 032109.
\bibitem{USCc} A. G\"{u}nter, A. A. Anappara, J. Hees, A. Sell, G. Biasiol, L. Sorba, S. De Liberato, C. Ciuti, A. Tredicucci, A. Leitenstorfer, and R. Huber, Nature 458 (2009) 178.
\bibitem{USCd} T. Niemczyk, F. Deppe, H. Huebl, E. P. Menzel, F. Hocke, M. J. Schwarz, J. J. Garcia-Ripoll, D. Zueco, T. H\"{u}mmer, E. Solano, A. Marx, and R. Gross, Nature Phys. 6 (2010) 772.
\bibitem{USCe} P. Forn-Diaz, J. Lisenfeld, D. Marcos, J. J. Garcia-Ripoll, E. Solano, C. J. P. M. Harmans, and J. E. Mooij, Phys. Rev. Lett. 105 (2010) 237001.

\bibitem{DSCa} J. Casanova, G. Romero, I. Lizuain, J. J. Garcia-Ripoll, and E. Solano, Phys. Rev. Lett.  105 (2010) 263603.
\bibitem{DSCb}S. Agarwal, S. M. Hashemi Rafsanjani and J. H. Eberly, J. Phys. B: At. Mol. Opt. Phys. 46 (2013) 224017.

\bibitem{Longhia} S. Longhi, Opt. Lett. 36 (2011) 3407.
\bibitem{Longhib} A. Crespi, S. Longhi, and R. Osellame, Phys. Rev. Lett. 108 (2012) 163601.

\bibitem{Rodriguez} B. M. Rodr\'{i}guez-Lara, F. Soto-Eguibar, A. Z. C\'{a}rdenas, and H. M. Moya-Cessa, Opt. Express 21 (2013) 12888.

\bibitem{Braaka} D. Braak, Phys. Rev. Lett. 107 (2011) 100401.
\bibitem{Braakb} E. Solano, Physics 4 (2011) 68.
\bibitem{Braakc} H. Zhong, Q. Xie, M. T. Batchelor, and C. Lee, J. Phys. A Math. Theor. 46 (2013) 415302.


\bibitem{add3} 
R. G. Brewer and E.L. Hahn, Phys. Rev. A11 (1975) 1641.

\bibitem{add4}
 B. Sobolewska, Opt. Commun. 19 (1976) 185.
 
\bibitem{add5} 
 R.M. Whitley and C.R. Stroud Jr, Phys. Rev. A 14 (1976) 1498.
 
\bibitem{add7}
C.M. Bowden and C.C. Sung, Phys. Rev. A 18 (1978) 1588.
\bibitem{add10} 
R.J. Cook and H.J. Kimble, Phys. Rev. Lett. 54 (1985) 1023.





\bibitem{NstatesRabia} C.C. Gerry and J. H. Eberly, Phys. Rev. A 42 (1990) 6805.
\bibitem{NstatesRabib} M. Alexanian and S.K. Bose, Phys. Rev. A  52 (1995) 2218.
\bibitem{NstatesRabic} V.V. Albert, Phys. Rev. Lett. 108 (2012) 180401.


\bibitem{uff1}
T.-S. Ho and S.-I. Chu, Phys. Rev. A 31 (1985) 659.
\bibitem{uff2}
T.C. Kavanaugh and R.J. Silbey, J. Chem. Phys. 98 (1993) 9444.
\bibitem{uff3}
V. G. Arkhipkin, JETP 81 (1995) 24.
\bibitem{uff4}
S. Guerin, R.G. Unanyan, L.P. Yatsenko, and H.R. Jauslin, Opt. Express 4 (1999) 84.
\bibitem{uff0}
J.H. Shirley, Phys. Rev. 138 (1965) B979.
\bibitem{driven2}
Q. Xie, J. Phys. B: At. Mol. Opt. Phys. 42 (2009) 105501.

\bibitem{Longhi-review} S. Longhi, Laser Photon. Rev. 3  (2009) 243.

\bibitem{Garanovich}
I.L. Garanovich, S. Longhi, A.A. Sukhorukov, and Y.S. Kivshar, Phys. Rep. 518 (2012) 1
\end{thebibliography}
\end{document}